\documentclass[12pt]{article}

\usepackage{epsfig}
\usepackage{amsmath}
\usepackage{soul,color}
\usepackage{bm}
\usepackage{epsfig}
\usepackage{natbib}
\usepackage{setspace}

\def\b#1{\mbox{\boldmath $#1$}}    

\newcommand{\pa}{\partial}         
\renewcommand{\th}{\theta}

\newcommand{\al}{\alpha}
\newcommand{\be}{\beta}
\newcommand{\de}{\delta}

\newcommand{\si}{\sigma}

\def\baselinestretch{1.8}
\usepackage{bm}
\usepackage{epsfig}

\begin{document}

\title{Mixtures of equispaced normal distributions and their use
for testing symmetry in univariate data}
\author{Silvia Bacci\footnote{Department of Economics, Finance and Statistics,
University of Perugia, Via A. Pascoli, 20, 06123 Perugia.} 
\footnote{{\em email}: silvia.bacci@stat.unipg.it} ,
Francesco Bartolucci$^*$\footnote{{\em email}: bart@stat.unipg.it}}   \maketitle
\def\baselinestretch{1.3}
\singlespacing

\def\baselinestretch{1.2}
\begin{abstract}\noindent
Given a random sample of observations, mixtures of normal densities are often used to 
estimate the unknown continuous  distribution from which the data come.  
Here we propose the use of this semiparametric framework for testing symmetry about an unknown value. 
More precisely, we show how the null hypothesis of symmetry may be formulated in terms of normal 
mixture model, with weights about the centre of symmetry constrained to be equal one another.  
The resulting model is nested in a more general unconstrained one, with same number of mixture 
components and free weights. Therefore, after having maximised the constrained and unconstrained 
log-likelihoods by means of  a suitable algorithm, such as the Expectation-Maximisation, symmetry is 
tested against skewness through a likelihood ratio statistic. The performance of the proposed 
mixture-based test is illustrated through a Monte Carlo simulation study, where we compare two 
versions of 
the test, based on different criteria to select the number of mixture components, with the 
traditional one based on the third standardised moment. An illustrative example is also given that 
focuses on real data. 

\end{abstract}\newpage

\def\baselinestretch{1.3}
\section{Introduction}

Let $X_1, X_2, \ldots, X_n$ be a random sample from a continuous distribution $F(x)$ with density 
$f(x)$. A problem which may be useful to consider is the symmetry of $f(x)$ about some unknown value. 
Indeed, nonparametric methods assume the symmetry of the distribution rather than its normality and, 
moreover, many parametric statistical methods are robust to the violation of the normality assumption 
of $f(x)$, being the symmetry often sufficient for their validity. For instance, in the context of 
regression models, \cite{bick:82} shows that, if the conditional density of errors is symmetric about 
zero, then the regression coefficients may be estimated in an adaptive way. Knowledge about the 
symmetry of $f(x)$ is also relevant to choose which location parameter is more representative of the 
distribution, being mean, median, and mode not coincident in case of skewness.  Another situation in 
which testing for symmetry may be important is encountered in case-control studies, which
require the 
exchangeability of the joint distribution of observations of treated and controlled individuals. As 
exchangeability implies the symmetry of the distribution, knowing that a distribution is skewed 
allows to exclude its exchangeability \citep{hol:88}.

By indicating with $\mu$ the mean or the median of $f(.)$,  the problem of testing symmetry may be formulated as  
\[
H_0: \quad F(\mu-x) = 1- F(\mu+x) \quad \forall x
\]
against the alternative hypothesis of skewness
\[
H_1: \quad F(\mu-x) \neq 1- F(\mu+x)
\]
for at least one $x$. Several procedures have been proposed in the literature to solve this 
testing problem
(for a review see \cite{hol:06}) and they can be classified on the basis of the used skewness 
measurement.  
 
The most known skewness index is given by the third standardised moment $\gamma_1$, usually estimated 
by the corresponding sample moment $b_1$. \cite{gupta:67} proves 
that, provided that $F(x)$ has finite 
central moments up to order six, $b_1$ is asymptotically normally distributed with a variance well 
defined. By estimating this variance through  the corresponding sample moments, an asymptotically 
distribution-free test is obtained. Another test based on $b_1$ is presented by  \cite{Dag:70}, 
who proposes a suitable transformation of $b_1$ having standard normal distribution already for 
small sample sizes (see also \cite{dag:pea:73} and
\cite{dag:90} for more details). This test is quite popular 
being one of the few symmetry tests implemented in widespread statistical packages, such as Stata. 
However, the main drawback is that it assumes the normality of $F(x)$ 
under the null hypothesis, so 
ignoring the possible presence of excess or defect of kurtosis in the distribution.   
 
Although $\gamma_1$ is a traditional measure of skewness, it is not free of drawbacks:  it is 
sensitive to outliers and it can even be undefined for heavy-tailed distributions such as the Cauchy; 
moreover, although it is equal to zero for symmetric distributions, a value of zero does not 
necessarily mean that the distribution is symmetric \citep{ord:68, joh:kotz:70}. Therefore, 
other symmetry tests have been developed which are based on alternative measures of skewness, 
such as those proposed by \cite{yule:1911} and 
\cite{bon:1933}, that take into account the difference between mean and median of population. 
\cite{cab:mas:96}  propose a test based on an asymptotically normally distributed estimator of the 
Yule's skewness index. They obtain an asymptotic distribution-free test by using the asymptotic 
variance derived under the normality assumption, discussing that the misspecification effect is 
negligible for the main part of practical problems. More recently, \cite{miao:gel:gas:06} modify the 
procedure of \cite{cab:mas:96}, by substituting the sample standard deviation with a function of 
differences (in absolute value) between each observation and the sample median,  and find a test that 
is generally more powerful. Finally, \cite{mira:99} proposes a test based on an estimator of the 
Bonferroni's index and provides consistent estimate  for the variance of the test-statistic.

Apart from the previously mentioned tests, 
several others exist which are based on different skewness measures. 
One of the most well-known 
is a nonparametric test proposed by \cite{gupta:67}, based on the concept of 
stochastic dominance, in which the test statistic is given by the  difference between the number of 
positive and negative (in absolute value) deviations from median. Instead, \cite{rand:fli:80}  
propose a triples test, in which  observation triples are considered and the presence of skewness is 
assessed by a suitable function of the difference between the number of right triples (i.e., when the 
middle observation in a given triple is closer to the smallest one) and that of left triples (i.e., 
when the middle observation is closer to the largest one). Another interesting class of tests is 
represented by the runs tests \citep{mcwill:90, mod:gas:96}: after having ordered the observations 
according to the absolute value and retaining signs, the number of changes of sign (so called runs) 
in the sequence gives an indication about the symmetry of the distribution. 

The problem of testing symmetry has also received considerable attention in these last years. Without 
pretending to be exhaustive, we remind the test of symmetry  of \cite{hol:10} based on a skewness 
index that combines the third standardised moment with a suitable function of the difference between 
mean and median; the proposal of \cite{zhe:gast:10}  of using the bootstrap method to estimate, in 
presence of small sample sizes, the distribution of test-statistic for some known tests; the solution 
of \cite{abd:but:11} for testing symmetry in presence of right censure. We also remind the works of 
\cite{ley:09} and \cite{cas:hal:pai:11} for the analysis of local optimality properties of some 
parametric, semi- and non-parametric tests, showing that the traditional test based on $b_1$ is 
optimal in proximity of normal distributions.

Finally, we observe that some part of the
recent literature has addressed the problem of testing symmetry 
by adopting the nonparametric kernel estimation method. Among others, \cite{fan:gen:95} propose, in 
the context of linear regression, a test for symmetric error distribution based on the kernel 
estimation of the density function of the errors; \cite{nga:06} illustrates several tests based on 
kernel estimators of skewness measures alternative to the traditional ones; \cite{rac:maa:07} 
describe a kernel-based test that uses a metric entropy statistic. The use of kernel estimators is an 
interesting one, because, being a nonparametric method, it allows a better goodness of fit with 
respect to parametric methods; on the other hand, it suffers from a high number of unknown 
parameters. An alternative approach to overcome this drawback and to which we focus in this 
contribution  is represented by the normal finite mixture (NM) models 
\citep{titt:smith:mark:85,lindsay:1996,mcla:peel:00}. Because 
any continuous - symmetric or skewed - distribution can be approximated arbitrarily well by a finite 
mixture of normal densities with common variance \citep{ferg:83},  NM provide a convenient 
semiparametric framework in which to model unknown distributional shapes, by keeping (\emph{i})  a 
parsimony close to that of full parametric methods as represented by a single  density  and (\emph 
{ii}) the flexibility of nonparametric methods as represented by the kernel method
\citep{esc:west:95,rob:96,roed:was:97}. 

Aim of the present
paper is to propose the use of NM for testing symmetry of a distribution about 
an unknown value. Indeed, already the first studies of \cite{pearson:1894} outline how a skewed 
distribution can be well described through a mixture of two normal densities. The general idea is 
that if the sample observations come from a symmetric distribution, then the weights of mixture 
components equidistant from the centre of symmetry are equal, being different otherwise. Therefore, 
we show how the above mentioned null hypothesis $H_0$ of symmetry can be alternatively expressed in 
terms of constraints on the weights. We also outline how a critical point in the proposed testing 
procedure concerns the choice of the number of mixture components: in particular, we base our 
choice on some well-known and commonly accepted information criteria, following  \cite{mcla:peel:00}. 
Then, to decide whether to reject or not
the null hypothesis, we illustrate that a likelihood ratio test 
may be performed by comparing, as usual, the maximum unconstrained log-likelihood of the model with 
the maximum constrained log-likelihood 
(i.e., under $H_0$).  More precisely, we show how to compute 
the log-likelihood function of the NM used for testing symmetry and we also show how to maximise it 
through an Expectation-Maximisation (EM) algorithm. 

The performance of the proposed approach is illustrated through a Monte Carlo simulation study that 
compares our proposed test with the traditional test  of \cite{gupta:67},
which is based on the third sample 
standardised moment. Our test is evaluated by selecting the optimal number of mixture components by 
using both Akaike's and Bayesian Information Criteria (AIC and BIC, respectively). Finally, an 
application to real data is illustrated.

The paper is organised as follows. In Section 2 we describe the main characteristics of an NM model 
and  we illustrate the EM algorithm  implemented to maximise the log-likelihood of the model. 
Moreover, the proposed test of symmetry is formulated in terms of constraints on the weights.
The main results of the simulation study are shown in 
Section 3, where our test is compared with the traditional test based of the third sample 
standardised moment, whereas in Section 4 we describe the application of our proposed test to real 
data. Finally, some remarks conclude our work.
\section{The mixture-based test of symmetry}
In the following, we describe the main characteristics of the NM model on which our test of symmetry 
is based and we give some indications about finding the maximum log-likelihood through an EM 
algorithm. Finally, we formulate the null hypothesis of symmetry as constraints about weights of NM 
model and we describe how to verify it by a likelihood ratio test.

\subsection{Mixture model}
Let $k$ be  the number of normal components of the mixture, 
let $\al$ be the centre of the symmetry, 
and let $\be$ be a scale parameter such that the support points of the mixture are
\[
\nu_j = \al + \be \de_j,\quad j = 1,\ldots,k,
\]
where $\de_1,\ldots,\de_k$ is a grid of equispaced points between
$-1$ and $1$. Therefore, the density of a mixture of $k$ normal components (NM$_k$)
results defined as
\[
f(x) = \sum_{j=1}^k \pi_j \phi(x;\nu_j,\si^2),
\]
where $\phi(x;\nu_j,\si^2)$ denotes the density at $x$ of the distribution $N(\nu_j,\si^2)$.

The model log-likelihood is given by
\[
\ell(\b\th) = \sum_{i=1}^n\log\sum_{j=1}^k \pi_j \phi(x_i;\nu_j,\si^2),
\]
where the parameters with respect to which it has to be maximised are $\b\th = (\al,\be,\pi_1,\ldots,\pi_k)$. To
compute these estimates, we can make use of the well-known EM
algorithm of \cite{demp:lair:rubi:77}, which is
described in detail in the following section.

\subsection{EM algorithm}
To introduce this algorithm consider the {\em complete data
log-likelihood}
\[
\ell_c(\b\th) = \sum_{i=1}^n\sum_{j=1}^k z_{ij}
\log\phi(x_i;\nu_j,\si^2)+\sum_j z_{\cdot j}\log\pi_j,
\]
where $z_{ij}$ is a dummy variable equal to $1$ if the $i$-th
observation belongs to the $j$-th component and to $0$ otherwise
and $z_{\cdot j} = \sum_i z_{ij}$. The EM algorithm is based on
the following two steps, to be performed until convergence:

\begin{itemize}
\item[(E)] compute the expected value of $z_{ij}$,
$i=1,\ldots,n$ and $j=1, \ldots,k$, given the observed data $\b x=(x_1,\ldots,x_n)$ 
and the current
value of the parameters $\b\th$; in practice this expected value
is computed as
\[
\hat{z}_{ij} = \frac{\phi(x_i;\nu_j,\si^2)\pi_j}{\sum_h
\phi(x_i;\nu_h,\si^2)\pi_h}.
\]
\item[(M)] maximise $\ell_c(\b\th)$ with any $z_{ij}$ substituted
by $\hat{z}_{ij}$. The derivatives of $\ell_c(\b\th)$
with respect to $\al$ and $\be$ are, respectively,
\begin{eqnarray*}
\frac{\pa \ell_c(\b\th)}{\pa\al} &=& \sum_i\sum_j
z_{ij}\frac{x_i-\nu_j}{\si^2},\\
\frac{\pa \ell_c(\b\th)}{\pa\be} &=& \sum_i\sum_j
z_{ij}\frac{x_i-\nu_j}{\si^2}\de_j.
\end{eqnarray*}
So, after some algebra, we can easily see that the solution is
reached when:
\begin{eqnarray*}
\be &=& \frac{\sum_i\sum_j z_{ij}(x_i-\bar{x})\de_j}{\sum_j
z_{\cdot j}(\de_j-\bar{\de})\de_j},\\
\al &=& \bar{x}-\be\bar{\de},\\
\si^2 &=& \frac{\sum_i\sum_j z_{ij}[x_i-(\al+\be\de_j)]^2}{n},
\end{eqnarray*}
where $\bar{x} = \sum_i x_i/n$ and $\bar{\de} = \sum_j z_{\cdot
j}\de_j/k$. The maximisation with respect to the
parameters $\pi_j$'s is simply
\begin{equation}\label{eq:weights}
\hat{\pi}_j = \frac{\hat{z}_{\cdot j}}{n},\quad j=1,\ldots,k.
\end{equation}
\end{itemize}

A crucial point with NM models concerns the choice of the number $k$ of mixture 
components. When the main aim of adopting an NM model is to use a semiparametric  framework for 
density estimation, as in our case, rather than the clustering of observations, \cite{mcla:peel:00}
[Chap. 6] (see also references cited therein)  discuss that the well-known AIC \citep{aka:73} and BIC 
\citep{sch:78} indices present an adequate performance for choosing $k$. Coherently, we suggest to 
use these criteria, although they may lead to different choices of $k$. More precisely, AIC tends to 
overestimate the true number of components. Moreover, we only select $k$ as an odd number, so
that there is one mixture component, the $[(k+1)/2]$-th, which corresponds to the centre of
the distribution and its mean directly corresponds to the parameter $\al$.
\subsection{Proposed test of symmetry}
In the proposed NM framework, the hypothesis of symmetry may be 
formulated as
\[
H_0:\pi_j=\pi_{k-j+1},\quad j=1,\ldots,[k/2],
\]
where $[z]$ is the largest integer less or equal than $z$ and $k$ is fixed. In other words, in a 
symmetric density the components specular with respect to the centre of symmetry are represented in 
equal proportions, whereas in a skewed density they are mixed in different proportions. 

We observe that the NM$_k$ model under $H_0$ is nested in the NM$_k$ model
with unconstrained $\pi_j$. 
Therefore, for testing symmetry we may use a likelihood ratio test, based on the deviance
\[
dev = 2[\ell(\hat{\b\th}) -  \ell(\hat{\b\th}_0)],
\]
where $\hat{\b\th}$ is the unconstrained maximum likelihood estimator
of $\b\th$ and $\hat{\b\th}_0$ is that under the constraint $H_0$, obtained according to the above 
described EM algorithm. We note that, under $H_0$, the estimator  $\hat{\pi}_j$ in equation 
(\ref{eq:weights}) becomes
\[
\hat{\pi}_j = \frac{\hat{z}_{\cdot j}+\hat{z}_{\cdot k-j+1}}{2n},\quad j=1,\ldots,k.
\]

Under $H_0$, 
$dev$ is distributed as a Chi-square with a number of degrees of freedom equal to  $[k/2]$, 
that is the 
number of constrained weights. We observe that when $k=1$ is selected the NM degenerates to a single 
normal distribution and, therefore, the null hypothesis of symmetry results automatically accepted.

Note that, when $H_0$ is true, $k$ reflects the true number of latent groups in which the population 
units are clustered. Instead, in presence of a skewed density, $k$ depends both on the groups 
characterising the population and on the level of skewness, because more than one normal component 
is usually needed to model a skewed  distribution. Therefore, there is not any more a one-to-one 
correspondence between the mixture components and the groups.
\section{Monte Carlo study}
This section summarises the results of 
a Monte Carlo study which shows the performance of the proposed test of 
symmetry. Two versions based on AIC and BIC are compared with the traditional test based on the third 
sample standardised moment
\[
b_1 = \frac{m_3}{m_2^{3/2}},
\]
where $m_r$ is the sample central moment of order $r$, given by $m_r = 1/n \sum_{i=1}^n (x_i - \overline{x})^r$. 

As outlined in the Introduction, $b_1$ is commonly used to estimate the third
standardised population moment
\[
\gamma_1 = \frac{\mu_3}{\mu_2^{3/2}},
\]
with $\mu_r = E[(X - \mu)^r]$. For samples from a symmetric 
distribution with finite sixth order central moment, \cite{gupta:67} shows that $b_1$ is 
asymptotically normally distributed with mean equal to 0 and variance equal to
\[
\sigma^2 = \frac{\mu_6 - 6\mu_2\mu_4 + 9\mu_2^3}{n\mu_2^3},
\]
which may be consistently estimated by substituting $\mu_j$, $j = 2, 4, 6$, with the appropriate 
sample moments. Therefore, under the null hypothesis of symmetry, the test-statistic 
\[
S_1 = \frac{n^{1/2} b_1}{\hat\sigma}
\]
has asymptotic standard normal distribution.

Within the Monte Carlo study we simulated 1000 samples with increasing size 
($n = 20, 50, 100$) from the 
following distributions: a standard normal ($N(0, 1)$), a Student's $t$ with 5 degrees of freedom 
($t_5$), a Laplace or double exponential ($Lap$), a symmetric mixture of three normal distributions 
(NM$_3$), a Chi-square with 1 degree of freedom ($\chi_1^2$), a Chi-square with 5 degrees of freedom  
($\chi_5^2$), a Chi-square with 10 degrees of freedom ($\chi_{10}^2$), and a lognormal with 
mean 0 and variance equal to 1 ($logN$). All tests are performed for nominal levels $\alpha$ equal to 
$0.01, 0.05, 0.10$. All analyses are implemented in R software.

The comparison among the two versions of the proposed mixture-based test and the Gupta's $S_1$-based 
test is performed by taking into account the following optimality criteria for a good test: (\emph
{i}) the empirical type-I error probability must not be higher than the nominal significance level
for distributions satisfying the null hypothesis of symmetry and (\emph{ii}) the 
empirical power for skewed alternatives must be as better as possible. Both of these informations are 
included for the simulated data in Tables \ref{tab:sim1} and \ref{tab:sim2}, respectively. 

As concerns the empirical significance level (Table \ref{tab:sim1}), the mixture-based test 
shows a performance very similar to that of Gupta's test when the number $k$ of components is 
selected by means of BIC. On the contrary, when AIC is used for the model selection, an empirical 
level is observed constantly higher than the nominal one: in other words, the type-I error is 
committed too often. This may be explained through results in Table \ref{tab:kvalues1}, showing the 
empirical percentage frequencies distributions of optimal $k$ values selected according to AIC or 
BIC. In case of data from symmetric distributions, the $k$ value should coincide with the actual 
number of groups, that is one for the first three cases and three for the NM$_3$. However, as we can 
observe from Table \ref{tab:kvalues1}, the AIC method overestimates $k$ more often than
the BIC method.

\begin{table}[!ht]
\begin{center}
\begin{tabular}{llrrrr}
  \hline
  	&	\multicolumn1c{$n$}	&	\multicolumn1c{$N(0,1)$}	&	\multicolumn1c{$t_5$}	&	
	\multicolumn1c{$Lap$} 	&	\multicolumn1c{NM$_3$}	\\
  \hline\hline
 \multicolumn6c{$\alpha= 0.01$}  \\ 
Mixture test (AIC)	&	20	&	0.018	&	0.019	&	0.019	&	0.036	\\
	&	50	&	0.024	&	0.024	&	0.015	&	0.020	\\
	&	100	&	0.029	&	0.027	&	0.021	&	0.019	\\
Mixture test (BIC)	&	20	&	0.011	&	0.003	&	0.009	&	0.026	\\
	&	50	&	0.007	&	0.007	&	0.006	&	0.015	\\
	&	100	&	0.004	&	0.007	&	0.008	&	0.015	\\
Gupta's Test	&	20	&	0.001	&	0.001	&	0.004	&	0.004	\\
	&	50	&	0.002	&	0.006	&	0.005	&	0.003	\\
	&	100	&	0.006	&	0.002	&	0.007	&	0.008	\\
 \multicolumn6c{$\alpha= 0.05$}  \\
Mixture test (AIC)	&	20	&	0.059	&	0.061	&	0.069	&	0.093	\\
	&	50	&	0.069	&	0.076	&	0.075	&	0.079	\\
	&	100	&	0.078	&	0.083	&	0.096	&	0.060	\\
Mixture test (BIC)	&	20	&	0.019	&	0.012	&	0.030	&	0.062	\\
	&	50	&	0.010	&	0.014	&	0.031	&	0.058	\\
	&	100	&	0.005	&	0.027	&	0.047	&	0.048	\\
Gupta's Test	&	20	&	0.038	&	0.030	&	0.044	&	0.037	\\
	&	50	&	0.038	&	0.029	&	0.035	&	0.045	\\
	&	100	&	0.043	&	0.032	&	0.037	&	0.045	\\
 \multicolumn6c{$\alpha= 0.10$}  \\
Mixture test (AIC)	&	20	&	0.101	&	0.099	&	0.130	&	0.144	\\
	&	50	&	0.088	&	0.125	&	0.145	&	0.133	\\
	&	100	&	0.096	&	0.134	&	0.167	&	0.136	\\
Mixture test (BIC)	&	20	&	0.028	&	0.031	&	0.062	&	0.097	\\
	&	50	&	0.012	&	0.030	&	0.061	&	0.094	\\
	&	100	&	0.005	&	0.053	&	0.080	&	0.107	\\
Gupta's Test	&	20	&	0.095	&	0.083	&	0.114	&	0.089	\\
	&	50	&	0.083	&	0.077	&	0.104	&	0.093	\\
	&	100	&	0.083	&	0.083	&	0.094	&	0.092	\\
\hline
\end{tabular}
\caption{\em Empirical significance 
levels of the mixture test based on AIC, mixture test based on BIC, and Gupta's 
test at levels of significance $0.01, 0.05, 0.10$, based
on 1000 simulated sample of size $n = 20, 50, 100$ from certain 
symmetric distributions.}
\label{tab:sim1}
\end{center}
\end{table}

\begin{table}[!ht]
\begin{center}
\begin{tabular}{llrrrr}
  \hline
  	&	\multicolumn1c{$n$}	&	\multicolumn1c{$\chi_1^2$}	&	\multicolumn1c{$\chi_5^2$}	&	\multicolumn1c{$\chi_{10}^2$} 	&	\multicolumn1c{$logN$}	\\
  \hline\hline
   \multicolumn6c{$\alpha= 0.01$}  \\ 
Mixture test (AIC)	&	20	&	0.351	&	0.065	&	0.039	&	0.195	\\
	&	50	&	0.760	&	0.497	&	0.246	&	0.562	\\
	&	100	&	0.963	&	0.888	&	0.651	&	0.784	\\
Mixture test (BIC)	&	20	&	0.233	&	0.034	&	0.022	&	0.122	\\
	&	50	&	0.649	&	0.231	&	0.096	&	0.424	\\
	&	100	&	0.933	&	0.612	&	0.277	&	0.704	\\
Gupta's Test	&	20	&	0.075	&	0.017	&	0.007	&	0.050	\\
	&	50	&	0.199	&	0.132	&	0.077	&	0.123	\\
	&	100	&	0.330	&	0.447	&	0.304	&	0.145	\\
   \multicolumn6c{$\alpha= 0.05$}  \\
Mixture test (AIC)	&	20	&	0.566	&	0.229	&	0.140	&	0.421	\\
	&	50	&	0.868	&	0.700	&	0.457	&	0.712	\\
	&	100	&	0.984	&	0.949	&	0.787	&	0.878	\\
Mixture test (BIC)	&	20	&	0.422	&	0.115	&	0.059	&	0.305	\\
	&	50	&	0.825	&	0.335	&	0.147	&	0.649	\\
	&	100	&	0.968	&	0.690	&	0.326	&	0.834	\\
Gupta's Test	&	20	&	0.359	&	0.153	&	0.089	&	0.272	\\
	&	50	&	0.496	&	0.541	&	0.373	&	0.341	\\
	&	100	&	0.661	&	0.798	&	0.713	&	0.423	\\
   \multicolumn6c{$\alpha= 0.10$}  \\
Mixture test (AIC)	&	20	&	0.680	&	0.328	&	0.216	&	0.573	\\
	&	50	&	0.920	&	0.772	&	0.525	&	0.806	\\
	&	100	&	0.988	&	0.962	&	0.828	&	0.913	\\
Mixture test (BIC)	&	20	&	0.557	&	0.177	&	0.090	&	0.447	\\
	&	50	&	0.896	&	0.388	&	0.173	&	0.775	\\
	&	100	&	0.980	&	0.715	&	0.341	&	0.893	\\
Gupta's Test	&	20	&	0.533	&	0.336	&	0.212	&	0.447	\\
	&	50	&	0.686	&	0.754	&	0.617	&	0.537	\\
	&	100	&	0.835	&	0.917	&	0.876	&	0.605	\\
\hline
\end{tabular}
\caption{\em Empirical power levels of the mixture test based on AIC, 
mixture test based on BIC, and Gupta's 
test at levels of significance $0.01, 0.05, 0.10$, based on 1000 simulated
sample of size $n = 20, 50, 100$ from certain skewed 
distributions.}
\label{tab:sim2}
\end{center}
\end{table}

\begin{table}[!ht]
\begin{center}
\begin{tabular}{lrrrrrrrr}
  \hline
  		\multicolumn1c{$k$}	&	\multicolumn2c{$N(0,1)$}	&	\multicolumn2c{$t_5$}	&	\multicolumn2c{$Lap$} 	&	\multicolumn2c{NM$_3$}	\\
\hline
	  		&	\multicolumn1c{AIC}	&	\multicolumn1c{BIC}	& \multicolumn1c{AIC}	&	\multicolumn1c{BIC}	& \multicolumn1c{AIC}	&	\multicolumn1c{BIC} & \multicolumn1c{AIC}	&	\multicolumn1c{BIC}		\\
  \hline\hline
       \multicolumn9c{$n=20$}  \\ 
1	&	75.3	&	92.5	&	67.6	&	85.4	&	59.1	&	80.7	&	12.7	&	32.9	\\
3	&	19.5	&	6.9	&	25.3	&	13.4	&	30.9	&	17.3	&	69.8	&	61.0	\\
5	&	4.3	&	0.6	&	5.9	&	1.2	&	8.3	&	2.0	&	14.6	&	5.9	\\
$>$5	&	0.9	&	0.0	&	1.2	&	0.0	&	1.7	&	0.0	&	2.9	&	0.2	\\
      \multicolumn9c{$n=50$}  \\ 
1	&	80.6	&	98.0	&	56.6	&	83.5	&	38.1	&	73.6	&	0.2	&	3.4	\\
3	&	15.7	&	1.8	&	35.2	&	16	&	46.0	&	23.4	&	85.3	&	93.9	\\
5	&	3.3	&	0.2	&	6.9	&	0.4	&	13.9	&	2.9	&	11.2	&	2.7	\\
$>$5	&	0.4	&	0.0	&	1.3	&	0.1	&	2.0	&	0.1	&	3.3	&	0.0	\\
       \multicolumn9c{$n=100$}  \\ 
1	&	85.5	&	99.4	&	40.1	&	73.8	&	10.7	&	51.1	&	0.0	&	0.0	\\
3	&	11.4	&	0.6	&	46.3	&	25.2	&	53.8	&	44.0	&	89.0	&	98.8	\\
5	&	2.4	&	0.0	&	11.3	&	1.0	&	25.9	&	4.7	&	9.5	&	1.1	\\
$>$5	&	0.7	&	0.0	&	2.3	&	0.0	&	9.6	&	0.2	&	1.5	&	0.1	\\  
  \hline
\end{tabular}
\caption{\em Percentage frequencies of $k$ values selected by means of AIC and BIC
for 1000 samples of size $n = 20, 50, 100$ simulated from certain symmetric distributions.}
\label{tab:kvalues1}
\end{center}\vspace*{5mm}
\end{table}

\begin{table}[!ht]
\begin{center}
\begin{tabular}{lrrrrrrrr}
  \hline
  		\multicolumn1c{$k$}		&	\multicolumn2c{$\chi_1^2$}	&	\multicolumn2c{$\chi_5^2$}	&	\multicolumn2c{$\chi_{10}^2$} 	&	\multicolumn2c{$logN$}	\\
		\hline
	  		&	\multicolumn1c{AIC}	&	\multicolumn1c{BIC}	& \multicolumn1c{AIC}	&	\multicolumn1c{BIC}	& \multicolumn1c{AIC}	&	\multicolumn1c{BIC} & \multicolumn1c{AIC}	&	\multicolumn1c{BIC}		\\
  \hline\hline
     \multicolumn9c{$n=20$}  \\ 
 1	&	7.0	&	17.9	&	46.9	&	71.5	&	61.3	&	83.5	&	10.3	&	22.7	\\
3	&	30.4	&	40.2	&	38.6	&	24.7	&	29.3	&	15	&	46.2	&	50.5	\\
5	&	37.5	&	31.6	&	13.0	&	3.7	&	7.9	&	1.3	&	30.3	&	22.5	\\
$>$5	&	25.1	&	10.3	&	1.5	&	0.1	&	1.5	&	0.2	&	13.2	&	4.3	\\     
 \multicolumn9c{$n=50$}  \\ 
1	&	0.0	&	1.3	&	15.6	&	56.3	&	39.6	&	80.4	&	0.2	&	1.4	\\
3	&	13.5	&	25.7	&	40.6	&	36	&	39.5	&	18.2	&	23.5	&	41.3	\\
5	&	23.6	&	35.0	&	34.5	&	7.6	&	16.7	&	1.4	&	25.2	&	33.1	\\
$>$5	&	62.9	&	38.0	&	9.3	&	0.1	&	4.2	&	0.0 	&	51.1	&	24.2	\\            
                    \multicolumn9c{$n=100$}  \\
1	&	0.0	&	0.0 	&	0.8	&	24.7	&	12.8	&	63.1	&	0.0	&	0.0	\\
3	&	2.5	&	7.6	&	18.9	&	47.9	&	41.0	&	32.8	&	7.8	&	16.4	\\
5	&	6.3	&	15.0	&	47.5	&	24.1	&	33.8	&	4.0	&	10.0	&	23.9	\\
$>$5	&	91.2	&	77.4	&	32.8	&	3.3	&	12.4	&	0.1	&	82.2	&	59.7	\\
  \hline
\end{tabular}
\caption{\em Percentage frequencies of $k$ values selected by means of AIC and BIC for 
1000 samples of size $n = 20, 50, 100$ simulated from certain skewed distributions.}
\label{tab:kvalues2}
\end{center}
\end{table}

On the other hand, this tendency of the AIC method to choose a relatively high number of mixture 
components results in a good performance of the mixture-based test for skewed distributions. In this 
case, the empirical power is clearly better with respect to the variant using the
BIC method and, most of 
all, to the Gupta's test (Table \ref{tab:sim2}). We also  observe that the variant of mixture-based 
test using BIC is almost always more powerful than Gupta's test. The only exception is observed in 
correspondence of data generated from $\chi_{10}^2$ with $n=100$: this is  the case closest to 
symmetry among those considered in our study and, as shown in Table \ref{tab:kvalues2}, both AIC and 
BIC methods tend to select a small number of mixture components.  
For all the three types of test, we observe that, as the sample size increases, the empirical 
significance level remains constant and the empirical power increases.

In conclusion, the proposed test based on BIC method 
has a performance better or, at worst, similar to that of the traditional 
test of symmetry.
\section{Empirical example}
\begin{figure}[!ht]\centering\vspace{0.5cm}
\includegraphics[width=8cm]{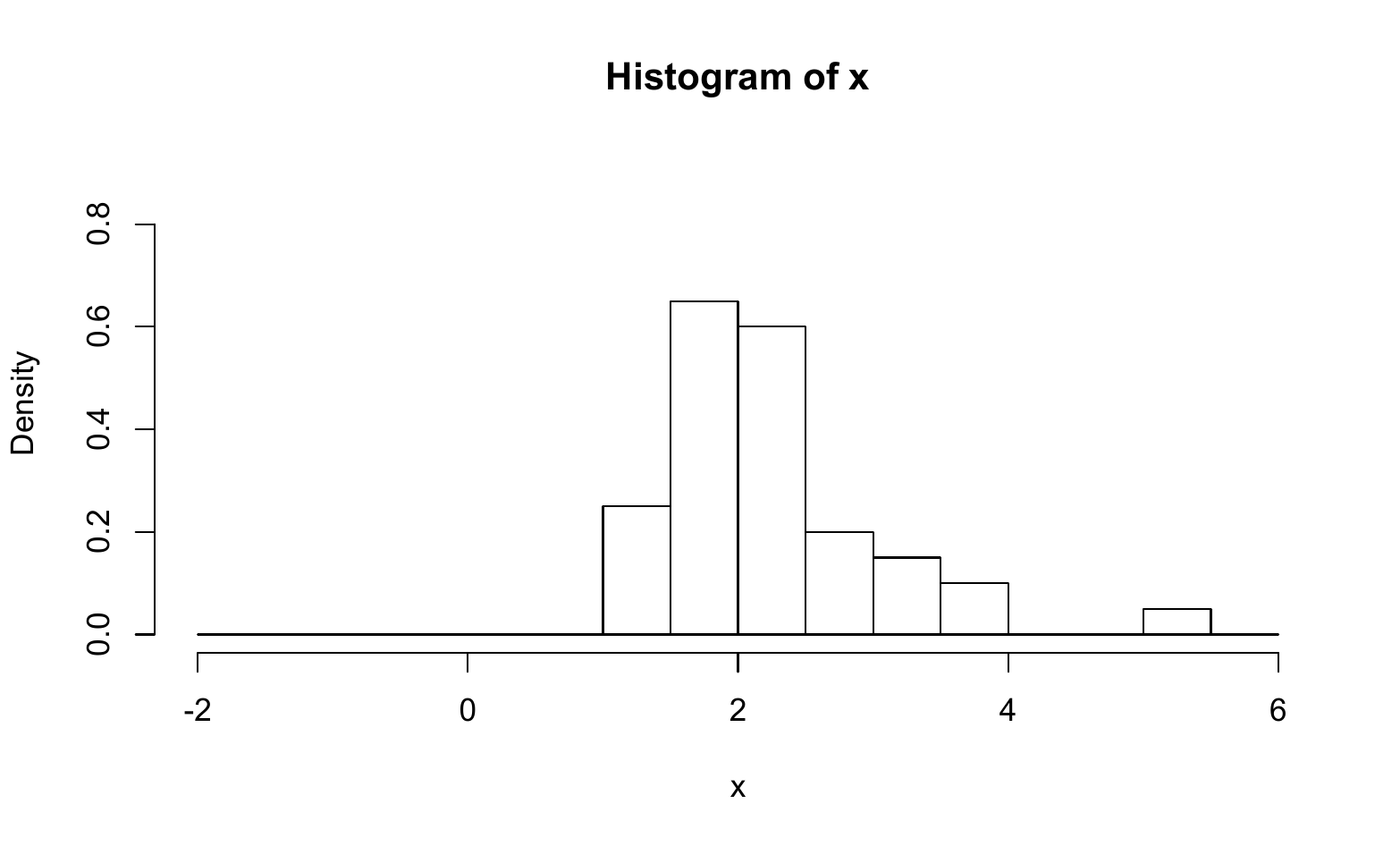}
\caption{\em Histogram of tomato roots data.}\vspace{0.5cm}
\label{fig:hist}
\end{figure}

In this section we illustrate the results obtained by testing the hypothesis
of symmetry for of a set of $n=40$ observations about tomato roots, whose histogram is 
represented in Figure \ref{fig:hist}. These data have been analysed through an NM model by 
\cite{gut:95} to identify the number of physical phenomena underlying the process of later root 
initiation. The
authors showed, through an approach based on the Box-Cox transformation, that the use of an NM$_2$ to 
adequately fit the data is due to the skewness of the data rather than to the presence of two 
physical phenomena behind the process at issue. Here we verify if their conclusions about skewness of 
data are confirmed by our test.

We first select the optimal number of mixture components by means of AIC and BIC. As shown in Table 
\ref{tab:tomato1}, for the general model with unconstrained weights, AIC 
index detects $k=5$ normal components, to which corresponds a log-likelihood equal to $-37.646$, 
whereas BIC index is more parsimonious and suggests to use $k=3$ components, to which corresponds a 
log-likelihood equal to $-40.554$. The corresponding AIC and BIC values ($89.292$ and 
$99.552$, respectively) are minimum also if we consider the case of the constrained model  (i.e. 
under $H_0$ true). In fact, in this last case the minimum AIC is given by $94.789$ and it is obtained 
for $k=3$, whereas the minimum BIC is equal to $101.545$, being again observed for $k=3$.\vskip5mm

\begin{table}[!ht]
\begin{center}
\begin{tabular}{l|crrr|crrr}
\hline
  	&	\multicolumn4{c|}{$H_0$ false}		&	\multicolumn4{c}{$H_0$ true}	\\
\hline
\multicolumn1{c|}{$k$}    &    \multicolumn1c{$\#$ par} &  \multicolumn1c{$\hat\ell$}    &  \multicolumn1c{AIC}    &   \multicolumn1{c|}{BIC}    &    \multicolumn1c{$\#$ par} & \multicolumn1c{$\hat\ell$} &     \multicolumn1c{AIC}    &   \multicolumn1c{BIC} \\
\hline \hline
1   &     2  & -47.583 &    99.165    &   102.543   &    2 & -47.583    & 99.165    &   102.543  \\
3   &     5 & -40.554  &  91.108   &   \textbf{99.552}    &  4 &  -43.394   &  94.789   &   101.545\\ 
5   &  7    &  -37.646  & \textbf{89.292}    &    101.114   &     5       & -42.558     &  95.116     & 103.560      \\      
7   &    9    &   -37.847&    93.695  &   108.900   &  6    &   -42.757   &   97.513   &   107.646  \\
\hline		
\end{tabular}
\caption{\em Number of mixture components selection: number of parameters, log-likelihood, AIC value, and BIC value under skewness and symmetry assumptions (in bold the minimum of AIC and BIC).}
\label{tab:tomato1}
\end{center}
\end{table}

In Table \ref{tab:tomato2} results of the deviance test based on the NM$_3$ and NM$_5$ model are 
shown. For $k=5$, with a $p$-value equal to $ 0.00736$ the null hypothesis of symmetry or, 
equivalently, the hypothesis that $\pi_1 = \pi_5$ and $\pi_2 = \pi_4$,
is strongly rejected in favour of that of skewness (i.e., at 
least one equality is not true). The same conclusion is reached by adopting $k=3$ mixture components, 
although the $p$-value is higher ($0.01715$). Note that the Gupta's test gives   $S_1 = 1.782$ with 
$p = 0.0748$, so leading to not reject the symmetry hypothesis.

\begin{table}[!ht]\vspace*{5mm}
\begin{center}
\begin{tabular}{lrr}
\hline
  	&		\multicolumn1c{$k=3$}	& \multicolumn1c{$k=5$} \\
\hline \hline
deviance      &   5.681   &  9.823 \\
df        &   1    &  2  \\
$p$-value    &  0.01715    &  0.00736 \\
\hline		
\end{tabular}
\caption{\em Mixture-based test for $k= 3, 5$: deviance, degrees of freedom, $p$-value.}
\label{tab:tomato2}
\end{center}
\end{table}

To conclude, analysed data may be described with a mixture of three or five normal components 
(according to the adopted model selection criterion). 
In both cases (Table \ref{tab:tomato3}) the main part 
of data is clustered in the second component ($\hat\pi_2 =  0.8804$, $\hat\mu_2=2.0154$ for $k=3$ and 
$\hat\pi_2=0.7569$, $\hat\mu_2=1.8863$ for $k=5$), followed by the third one ($\hat\pi_3=0.1196$, $
\hat\mu_3=3.9515$ and $\hat\pi_3= 0.1578$, $\hat\mu_3=2.9067$, respectively); very low is the 
representativeness of the first component. Variance is assumed to be constant over all the normal 
components, resulting equal to $0.2372$ for $k=3$ and to $0.1119$ for $k=5$.  
Finally, for $k=5$, components four and five gather respectively  $6.02\%$ and $2.51\%$ of 
observations with high average values.\vspace*{5mm}

\begin{table}[!ht]
\begin{center}
\begin{tabular}{lrr}
\hline 
  	&		\multicolumn1c{$k=3$}	& \multicolumn1c{$k=5$} \\
\hline \hline
$\hat\pi_1$    &    0.0000   &  0.0000  \\
$\hat\pi_2$    &  0.8804     &   0.7569  \\
$\hat\pi_3$    &   0.1196    &   0.1578 \\
$\hat\pi_4$    &  --      & 0.0602   \\
$\hat\pi_5$    &   --    & 0.0251   \\
$\hat\al$    &     2.0155  &    2.9067\\
$\hat\be$    &   1.9360    &   2.0407 \\
$\hat\mu_1$    &  0.0794     & 0.8660   \\
$\hat\mu_2$    &  2.0154     &  1.8863  \\
$\hat\mu_3$    &  3.9515     &  2.9067  \\
$\hat\mu_4$    &    --   &  3.9271  \\
$\hat\mu_5$    &   --    &  4.9474  \\
$\hat\sigma^2$    &    0.2372   &    0.1119 \\
\hline		
\end{tabular}
\caption{\em Parameter estimates under models NM$_3$ and NM$_5$.}
\label{tab:tomato3}
\end{center}
\end{table}

\begin{figure}[!ht]\centering\vspace*{0.5cm}
\begin{tabular}{cc} 
\includegraphics[width=8cm]{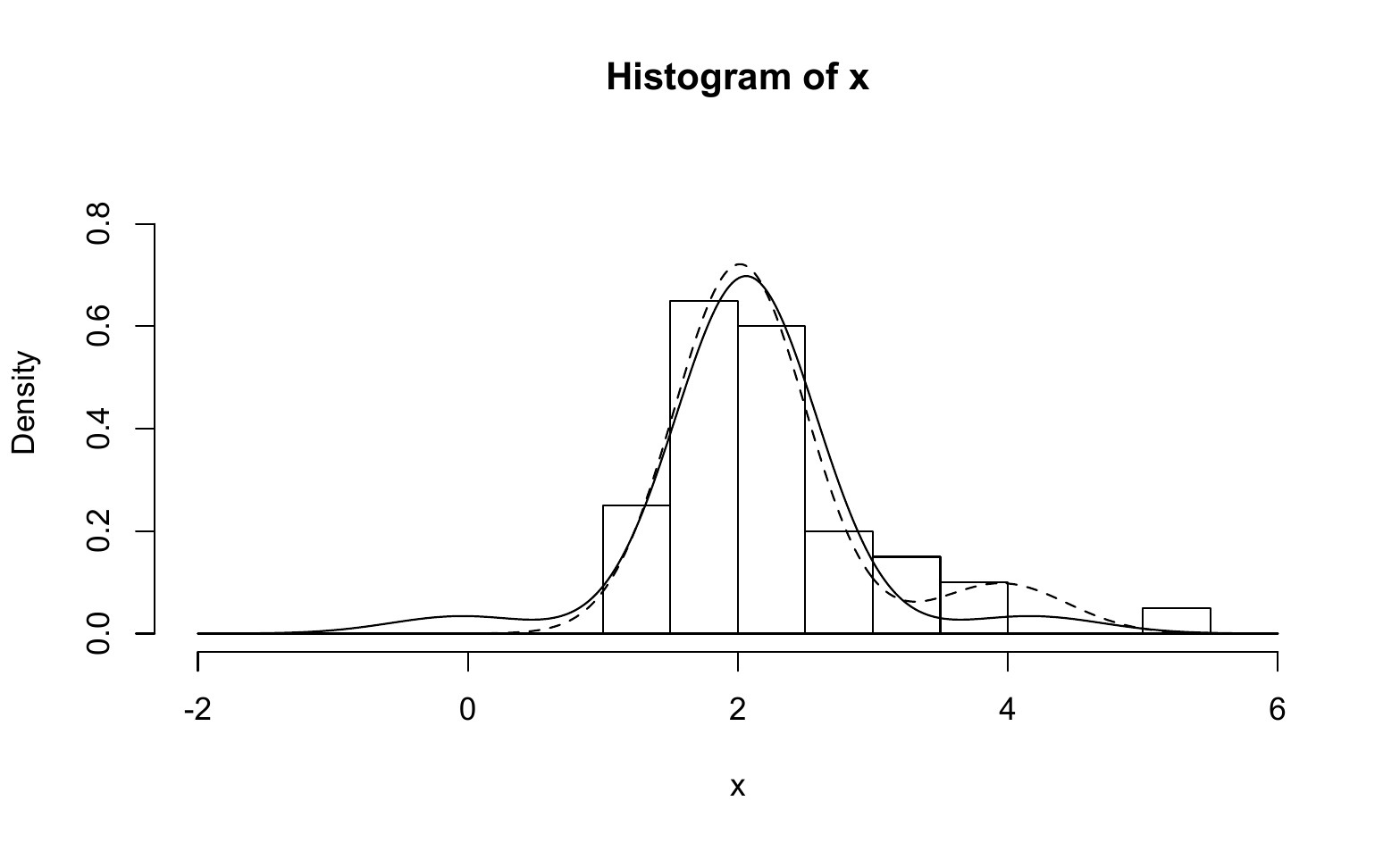} & 
\includegraphics[width=8cm]{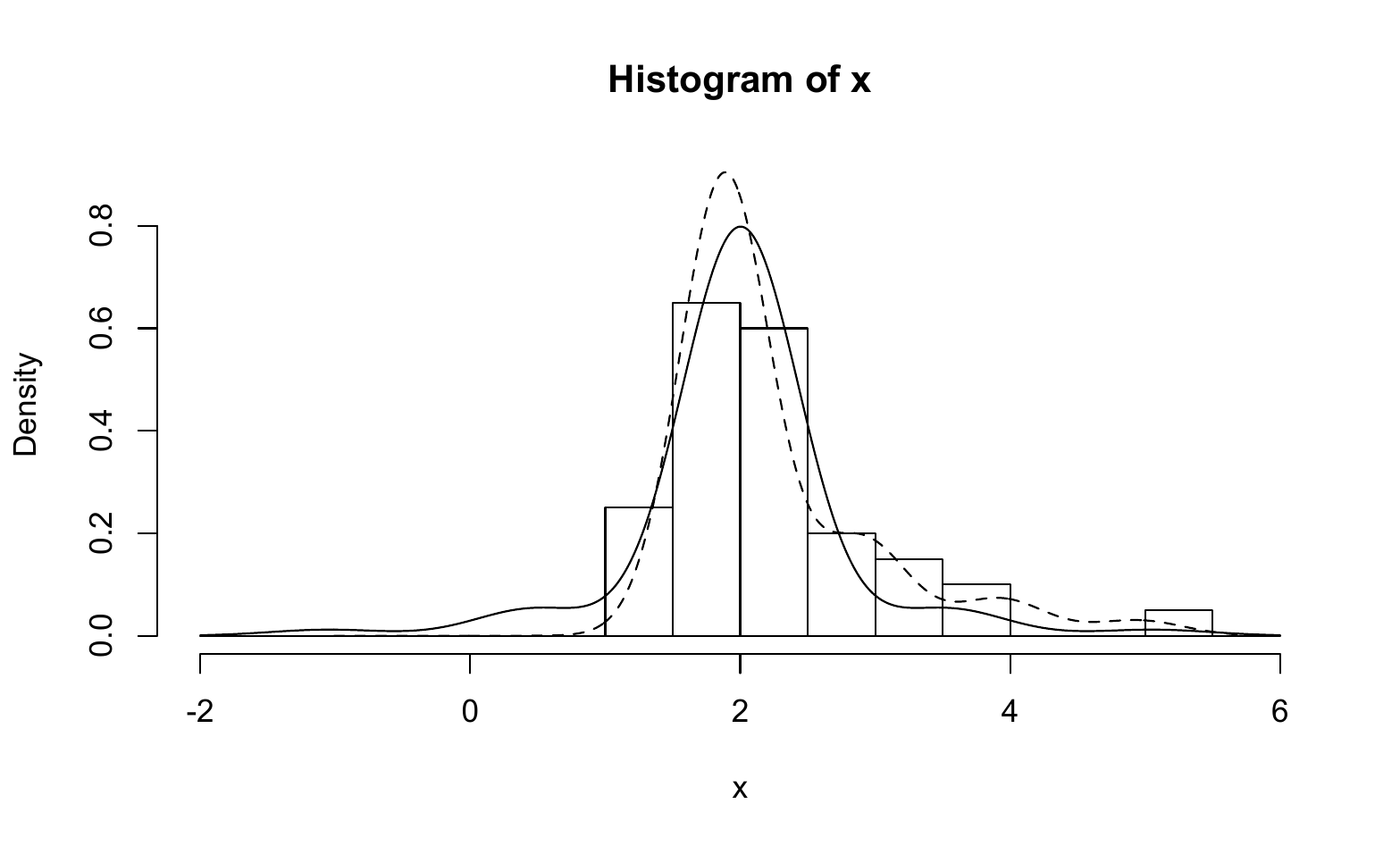}\\
$NM_3$ & $NM_5$
\end{tabular}
\caption{\em Histogram of tomato roots data with the estimated 
density under the unconstrained (dashed line) and constrained (solid line) NM$_3$  
and NM$_5$ models.}
\label{fig:hist2}\vspace*{0.5cm}
\end{figure}

\newpage Finally, in Figure \ref{fig:hist2} we show the estimated density under the
constrained and unconstrained NM$_3$ models (left 
panel) and NM$_5$ models (right panel)
overlapped to the histogram for the observed data. For both 
values of $k$, it can be clearly observed the better goodness of fit of the unconstrained NM model 
with respect to that constrained, allowing to take into account the positive skewness of data.

\section{Concluding remarks}
After having reviewed the literature concerning the issue of testing for symmetry, in this 
contribution we outlined the existence of an interesting framework  so far ignored, at least to our 
knowledge, to perform this test: that of normal mixture (NM) models. Indeed, NM models represent a 
semiparametric method to approximate unknown continuous densities   with a satisfying goodness of 
fit, most of all in presence of skewness. Therefore, they offer a natural setting in which to place 
the study of symmetry of a distribution.  

We first described the main characteristics of an NM model,
illustrating in detail the EM algorithm 
implemented for parameter estimation. Then, we formulated the hypothesis test at issue in 
terms of constraints on weights characterising the NM model.
Moreover, we describe how a likelihood ratio 
test is obtained, based on a test-statistic distributed according to a Chi-square with a number of 
degrees of freedom depending on the number of constraints and, therefore, on the number of mixture 
components.  

A Monte Carlo study outlined how the performance of the proposed test depends on the criterion used 
to select the number of mixture components. More precisely, we observed that using BIC a good 
empirical level of significance is obtained, comparable with that of the traditional test based on 
the  third standardised moment \citep{gupta:67}. On the other hand, the empirical power of our test
with BIC resulted usually better than that observed with Gupta's test. 

An analysis on real data about the process of later root initiation in tomatoes illustrated the 
application of the proposed mixture-based test. Both criteria used to select the number of mixture 
components allowed to conclude about the skewness of distribution, as opposed to Gupta's test, and to 
describe in detail the unknown underlying distribution from which data come. 
\bibliography{biblio}
\bibliographystyle{apalike}
\end{document}